\documentclass[12pt]{article}

\usepackage{lineno,hyperref}
\modulolinenumbers[5]
\usepackage{subfig}
\usepackage{amsmath,amsthm,amssymb,epsfig,bm,amsmath}
\usepackage{cleveref}
\usepackage{float}
\usepackage{amsmath}
\usepackage{textcomp} 
\usepackage{soul}
\usepackage{graphicx}
\usepackage{MnSymbol}
\usepackage{mathdots}
\usepackage{makecell}
\usepackage{listings}
\usepackage{caption}
\usepackage{amssymb}
\usepackage{subfig}   

\usepackage{footnote}
\usepackage[flushleft]{threeparttable}
\usepackage{multirow}
\usepackage{relsize}
\usepackage{lettrine}
\usepackage{courier}
\usepackage{color} 
\definecolor{mygreen}{RGB}{28,172,0} 
\definecolor{mylilas}{RGB}{170,55,241}
\definecolor{orange}{rgb}{.93,.53,.18}
\usepackage{amsmath,amsthm,amssymb,amsfonts} 
\usepackage[yyyymmdd]{datetime}
\usepackage{algorithm}
\usepackage[noend]{algpseudocode}

\usepackage{lipsum}
\usepackage{setspace}
\usepackage{siunitx}
\usepackage{pdfpages}
\usepackage{booktabs}
\usepackage[]{hyperref}
\hypersetup{
  colorlinks   = true, 
  urlcolor     = blue, 
  linkcolor    = black, 
  citecolor   = blue 
}
\usepackage[utf8]{inputenc}

\usepackage{soul} 
\soulregister\cite7
\soulregister\ref7
\soulregister\pageref7
 \graphicspath{{./figures/}}

\usepackage{listings}
\usepackage{color}
\definecolor{codegreen}{rgb}{0,0.6,0}
\definecolor{codegray}{rgb}{0.5,0.5,0.5}
\definecolor{codepurple}{rgb}{0.58,0,0.82}
\definecolor{backcolour}{rgb}{0.95,0.95,0.92}
 
\lstdefinestyle{mystyle}{
    backgroundcolor=\color{backcolour},   
    commentstyle=\color{codegreen},
    keywordstyle=\color{magenta},
    numberstyle=\tiny\color{codegray},
    stringstyle=\color{codepurple},
    basicstyle=\footnotesize,
    breakatwhitespace=false,         
    breaklines=true,                 
    captionpos=b,                    
    keepspaces=true,                 
    numbers=left,                    
    numbersep=5pt,                  
    showspaces=false,                
    showstringspaces=false,
    showtabs=false,                  
    tabsize=2
}
 
\usepackage{booktabs} 
\usepackage{siunitx} 

\usepackage{mathtools} 
\usepackage{gensymb} 

\usepackage{fancyhdr} 
\theoremstyle{definition}

\theoremstyle{definition}

\theoremstyle{remark}

\begin{document}

\title{\textbf{\Large Nuclear Reactor Safeguarding with \\ Neutrino Detection for MOX Loading Verification}}

\author{Bryan Helz$^{b,a}$, Leia Barrowes$^a$, Igor Jovanovic$^c$, \\  Dean Price$^c$, Brendan Kochunas$^c$, James D. Wells$^{a}$}

\date{\small
	{{$^a$}Department of Physics, University of Michigan,  Ann Arbor, MI 48109} \\
	{{$^b$}Department of Physics, University of Arizona, Tucson, Arizona 85721, USA} \\
	{$^c$Department of Nuclear Engineering and Radiological Science, University of Michigan, Ann Arbor, MI 48109}
}

\begin{spacing}{1.2}
	\maketitle
\end{spacing}
\vspace*{-0.5cm}
\begin{abstract}
The resurgence of interest in nuclear power around the world highlights the importance of effective methods to safeguard against nuclear proliferation.
Many powerful safeguarding techniques have been developed and are currently employed, but new approaches are needed to address proliferation challenges from emerging nuclear energy markets.
Building on prior work that demonstrated monitoring of nuclear reactor operation using neutrino detectors, we develop and present a simple quantitative statistical test suitable for analysis of measured reactor neutrino data and demonstrate its efficacy in a semi-cooperative reactor monitoring scenario.
In this approach, a moderate-sized neutrino detector is placed near the reactor site to help monitor possible MOX fuel diversion independent of inspection-based monitoring. 
We take advantage of differing time-dependent neutrino count rates during the operating cycle of a reactor core to monitor any deviations of measurements from expectations given a declared fuel composition. 
For a five-ton idealized detector placed 25m away from a hypothetical 3565 MW$_\text{th}$ reactor, the statistical test is capable of detecting the diversion of $\sim$ 80kg plutonium at the 95\% confidence level 90\% of the time over a 540-day observation period.
\end{abstract}

\centerline{\textit{Presented at the American Physical Society April Meeting, Minneapolis, 2023}}

\vfill\eject
\tableofcontents
\vfill\eject

\section{Introduction}

The idea of using neutrinos to monitor reactors is not new; in 1986, a neutrino laboratory was set up at the Rovno nuclear power plant in Kurchatov, Russia, and the reduction of 5\%--7\% in neutrino flux over burnup could be directly attributed to changes in the fuel composition~\cite{klimov1994neutrino}.
A case study~\cite{christensen2015antineutrino} suggested that antineutrino reactor monitoring with multiple detectors could have provided the IAEA the means to detect the illicit diversion of plutonium from a 5-MWe (mega-Watt electric) experimental reactor at Yongbyon, North Korea in 1994.

Recent experiments demonstrate the increasing practicality of neutrino detectors specifically for safeguards purposes.
The SONGS1 detector~\cite{bernstein2008monitoring}, installed at the San Onofre Nuclear Generating Station, was one of the earliest detectors used in an experiment for nonproliferation research. 
This detector, consisting of 0.64 tons of liquid scintillator and a muon veto system, was placed 25 meters away from the core of a 3.4-GWth reactor and demonstrated on-off detection with measurements recorded over seven-day intervals. 
As part of the MiniCHANDLER project~\cite{haghighat2020observation}, a small 80-kg detector was installed 25~m away from a 2.9-GWth reactor at the North Anna Nuclear Generating Station.
The experiment was considered successful as it was ``undertaken with the singular goal of demonstrating the detection of reactor neutrinos and their energy spectrum'', and it performed on-off detection for the reactor during the experimental period with a high level of confidence.
The PROSPECT~\cite{prospect} experiment and DANSSino~\cite{Alekseev_2014} detector achieved signal-to-noise ratios greater than one at around 10 meters from the reactor site.
The NUCIFER detector was designed to detect illicit Pu retrieval from a nuclear power plant~\cite{porta2010reactor, bouvet2012reactor, cucoanes2014nucifer}. 
It was designed to be portable, work within the safety regulations specified at nuclear reactor power plants, and be operable by non-neutrino experts.
The SANDD detector design was created to have directional sensitivity~\cite{sutanto2021sandd} to provide more effective background rejection for near-field reactor monitoring.
The next step is to understand in detail what would be needed to detect the diversion of a ``significant quantity" of nuclear material---defined to be 8~kg for $^{239}$Pu~\cite{atomowej2002iaea}.

In this study, we compute the time evolution of reactor neutrino flux and demonstrate a statistical test for detecting anomalous fuel loading in an operating reactor with a partially loaded mixed oxide (MOX) fuel core.
This tests a similar scenario to the one presented in Ref.~\cite{Bernstein_2018}, but improves upon its unidirectional MOX shift limitations and takes advantage of the different time-dependences of neutrino output from different reactor cores. 
The viability of the methodology employed in this study is evaluated under both systematic uncertainties which may arise from neutrino emission models as well as various sources of statistical uncertainty.

This paper contributes to the study of neutrino-detection methodology for the determination of conformity to safeguards agreements that permit the use of MOX fuel in PWRs, including small modular reactors \cite{nuscale2024}.
These safeguards agreements would provide strict guidelines on the use of MOX fuel in the reactor under surveillance.
The methodology accounts for various uncertainties associated with the detection process and provides a test statistic that can be used to determine the likelihood of a violation of the safeguards agreement.
We also present an application of this methodology to a hypothetical reactor whose neutrino output is calculated with simplified neutronics models.
The novelty of the method lies in its ability to provide a clear decision-making metric (the likelihood of a violation) based on measured data.

\section{Reactor Neutrino Model}

\subsection{Neutronics Modeling}
\label{sec:nmodel}

In this section we provide some details of our neutronics modeling that we will use in our simulations of neutrino flux. In particular, we need to understand the change in neutrino flux resulting from changes in the reactor's isotope-specific fission rates. 

In order to approximate these fission rates for an operational reactor, two pin cell models based on Ref.~\cite{kozlowski2006pwr} were created with the Serpent neutronics code.
Serpent is a continuous energy 3D Monte Carlo neutron transport code developed by VTT Technical Research Centre of Finland~\cite{leppanen2015serpent}.
The geometry of these pin cell models consists of a single fuel rod with reflective boundary conditions.
In reality, the PWR described in Ref.~\cite{kozlowski2006pwr} consists of 50,952 fuel pins that each have unique irradiation conditions and therefore isotopic compositions.
Furthermore, axial differences in moderator density and leakage will cause the composition of each fuel pin to vary axially as well.
The Serpent pin cell models represent this reality with a single fuel composition and burnup.

Table~\ref{tab:pin_cell_parameters} gives some of the modeling parameters used in the pin cell models.
One model is built to represent pins loaded with 4.3\% MOX fuel; the initial fissile content of the fuel consists of ${}^{235}$U and isotopes of Pu.
The MOX\% in fuel is defined as the percentage weight of $^{239}$Pu and $^{241}$Pu divided by weight of ${}^{239}$Pu, $^{240}$Pu, ${}^{241}$Pu, $^{242}$Pu, $^{234}$U, $^{235}$U, $^{236}$U and $^{238}$U.
The plutonium in a MOX pin is 93.6\% $^{239}$Pu, which means that this pin cell represents MOX fuel manufactured with weapons-grade plutonium.

The other pin cell model is built to represent pins loaded with pure UO$_2$  fuel with 4.2\% enrichment (4.2\% of the uranium mass is $^{235}$U), which qualifies it as a low-enriched uranium (LEU) fuel, a designation that we will often use below.
In this fuel, the entirety of the initial fissile content of the fuel is uranium.
The initial compositions of the fuels in these models are given in Table~\ref{tab:pin_cell_compositions}.
These were calculated based on a 4.5\% enrichment for the UO$_2$ pin cell and a 4.3 wt.\% MOX with natural (0.72\%) $^{235}$U enrichment for the MOX pin cell. 
Burnup calculations are run for these two models separately up to a maximum burnup of 50~GWD/tHMi.
``GWD/tHMi'' (gigawatt-day per metric ton of initial heavy metal) is a unit that describes the total energy output of the pin cell normalized by the initial mass of heavy metal in the fuel.
At each transport step, the Monte Carlo solution method is run until an uncertainty in $k_\mathrm{eff}$ of $5\times10^{-5}$ is reported, providing adequate spectral resolution of the neutron flux for its use in calculating isotope-specific fission rates. Reaching this uncertainty takes approximately $8 \times 10^7$ total histories. 
For comparison, a study with similar quantities of interest used MCNPX to perform burnup calculations on a pin cell model and used $1.5 \times 10^6$ total histories per burnup step~\cite{fensin2008improved}.

\begin{table}[t!]
\centering
\caption{Modeling parameters used in Serpent 2 pin cell models.}
\label{tab:pin_cell_parameters}
\begin{tabular}{ll}
\hline
Parameter & Value \\ \hline
Fuel pellet radius & 0.790 cm \\
Fuel cladding inner diameter & 0.802 cm \\
Fuel cladding outer diameter & 0.917 cm \\
Pin cell pitch & 1.26 cm \\
Gap material & Void \\
Cladding material & Zircaloy-2 \\
Coolant density & 0.712 g/cm$^3$ \\
Coolant Temperature & 600 K \\
Cladding Temperature & 600 K \\
Fuel Temperature & 900 K \\
Cross-section library & ENDF/B-VIII.0 \\ \hline
\end{tabular}
\end{table}

\begin{table}[t!]
\centering
\caption{Fuel compositions in UO$_2$ and MOX pin cell models.}
\label{tab:pin_cell_compositions}
\begin{tabular}{cccc}
\hline
\multicolumn{2}{c}{UO$_2$ Pin Cell} & \multicolumn{2}{c}{MOX Pin Cell} \\ \hline
Isotope & At. Density [at/b-cm] & Isotope & At. Density [at/b-cm] \\ \hline
${}^{16}$O & 0.04655 & ${}^{16}$O & 0.04642 \\
${}^{234}$U & 1.062 $\times 10^{-5}$ & ${}^{234}$U & 4.507 $\times 10^{-7}$ \\
${}^{235}$U & 1.058 $\times 10^{-3}$ & ${}^{235}$U & 4.487 $\times 10^{-5}$ \\
${}^{236}$U & 5.267 $\times 10^{-6}$ & ${}^{236}$U & 2.234 $\times 10^{-7}$ \\
${}^{238}$U & 0.02215 & ${}^{238}$U & 0.02211 \\
 &  & ${}^{239}$Pu & 9.898 $\times 10^{-4}$ \\
 &  & ${}^{240}$Pu & 6.213 $\times 10^{-5}$ \\
 &  & ${}^{241}$Pu & 4.195 $\times 10^{-6}$ \\
 &  & ${}^{242}$Pu & 1.044 $\times 10^{-6}$ \\ \hline
\end{tabular}
\end{table}

In this study, the ``MOX pin fraction'' (MPF) refers to the fraction of pins in a hypothetical reactor that would burn MOX fuel, and the remaining pins would burn LEU fuel. 
For example, the isotope-specific fission rates for a reactor with a 75\% MPF would be calculated by adding the isotope-specific fission rates from the MOX pin cell with a weight of 0.75, plus the isotope-specific fission rates from the UO$_2$ pin cell with a weight of 0.25.
The fission rates from this weighted sum are then scaled evenly to ensure the power output matches the 3565~MW$_\mathrm{th}$ reactor power given in Ref.~\cite{kozlowski2006pwr}.
This method includes some notable implicit assumptions---most importantly, the MPF does not necessarily equal the mass fraction of MOX fuel in the reactor, but rather the fraction of power derived from MOX fuel in the reactor.
In addition, this method also assumes each fuel rod has the same burnup over time.
These two assumptions are technically at odds with each other because of the difference in the initial heavy metal mass in the fuel of each pin cell.
Since the burnup unit is normalized by this mass, the difference causes the same power output for a fixed period of time to produce two different final burnup values.
Additional drawbacks of this pin cell-based modeling procedure include an inability to account for the effects of nonuniform core pin powers, control rods, burnable absorbers, or heterogeneous fuel loading that homogenization-based techniques~\cite{smith1986assembly} or state-of-the-art methods~\cite{kochunas2017vera} would account for.
Despite these assumptions, the isotope-specific fission rates generated from these models are used here to demonstrate the statistical tests described in Sect.~\ref{sec:tsf} as they still capture the dominant important aspects of the reactor behavior.
For a set fuel composition, since the total fission rate is approximately constant, only the neutron spectrum determines the isotope-specific fission rates.
These pin cell models account for many of the spectral phenomena observed in operating reactors (resonance self-shielding, neutron thermalization, fast fission, \textit{etc.}).
They also capture some burnup-related phenomena such as the buildup of fission products and heavy actinides and the depletion of $^{235}$U.
Furthermore, some uncertainty is introduced to these isotope-specific fission rates to account for uncertainties in cross-section data.
This is discussed further in Sect.~\ref{sec:tsup}.

\begin{figure}[t]
  \centering
    \includegraphics[width=0.49\textwidth]{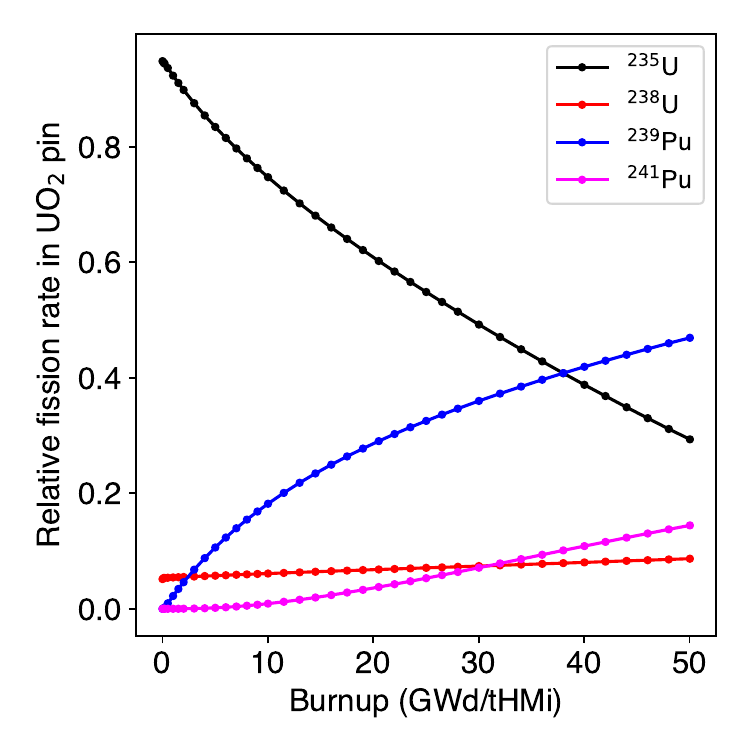}
    \includegraphics[width=0.49\textwidth]{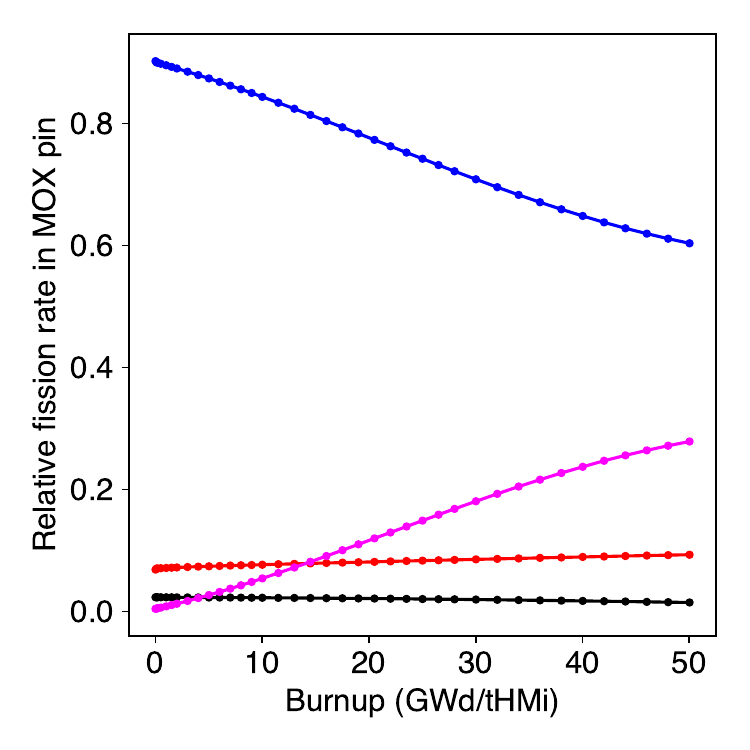}
    \caption{Left: Isotope-specific fractional fission rates for a UO$_2$ (synonymous with LEU) pin cell, which is equivalent to an MPF of zero.
    Right: Isotope-specific fractional fission rates for a MOX pin cell, which is equivalent to an MPF of 100\%.}
      \label{fig:relfiss}
\end{figure}

The relative fission rates for each of the four fissile isotopes for both the UO$_2$ and MOX pin cell models are given in Fig.~\ref{fig:relfiss}.
These fission rates are used to generate the neutrino rates expected in a detector.

\subsection{Neutrino Spectral Models}
\label{sec:nspec}
The next step in determining the reactor neutrino flux from a hypothetical reactor is to model neutrino spectrum for each of the four primary fissioning isotopes ($^{235}$U, $^{238}$U, $^{239}$Pu, and $^{241}$Pu).
There are two methods of doing so: the inversion method and the summation method.
The inversion method uses the measured electron spectrum of the fissioning nucleus and uses Fermi's theory of beta decay to calculate the neutrino spectra.
This method is limited by the irradiation time of the sample and the precision of the model of beta decay used in the inversion, \textit{i.e.,} whether or not effects such as the emission of photons, weak magnetism, and shape factors are considered in the inversion.
The summation method uses data from nuclear databases to sum all of the neutrino contributions from the relevant beta decays weighted by their activity.
It explicitly accounts for each branch of beta decay but is limited by missing data. 
The last decade's standard has been the Huber-Mueller (HM) model, which uses the inversion method for $^{235}$U, $^{239}$Pu, and $^{241}$Pu~\cite{huber2011determination}, and the summation method for $^{238}$U~\cite{mueller2011improved}. 
The result is the number of neutrinos released per MeV per fission by the four primary fissioning isotopes in the interval 2--8~MeV in bins with a width of 0.25~MeV.

Work has been done to improve methods of determining neutrino spectra.
Notably, the Estienne-Fallot model (EF)~\cite{Estienne_2019}  assimilated the newly collected nuclear data into the summation framework established in the 1950s~\cite{perkins_king_1958} and continuously improved as more data was made available.

A recent new summation technique, which we call the PRL2023 model~\cite{neutrino2022}, simulates beta-transitions with the phenomenological Gamow-Teller decay model, which improves fit to the reactor data.
Unless otherwise stated we use the PRL2023 model for our neutrino spectral model, including its uncorrelated uncertainties in the neutrino spectrum. 
Although different models predict different neutrino emission rates, the results are consistent and produce very similar results for the resolving power of the statistical test that we apply.
Additionally, across the smallest changes in MPF considered in this study, the difference in neutrino spectra due to the changes in MPF is far greater than differences resulting from uncertainty in the neutrino spectral models.

\subsection{Background Counts}


\noindent
{\it Antineutrino background}

Geoneutrinos and diffuse supernova background comprise non-reactor antineutrino sources in the 2--8~MeV detection energy range~\cite{Vitagliano_2020}.
Geoneutrinos originate from radioactive decay in the Earth.
Their flux varies geographically within an order of magnitude depending on crustal thickness, radioactivity, and surface heat flux~\cite{Smirnov_2019}.
Diffuse supernova neutrino flux has a more constant energy spectrum over the energy range of interest.

\begin{figure}[t]
    \centering
    \includegraphics[width=0.6\textwidth]{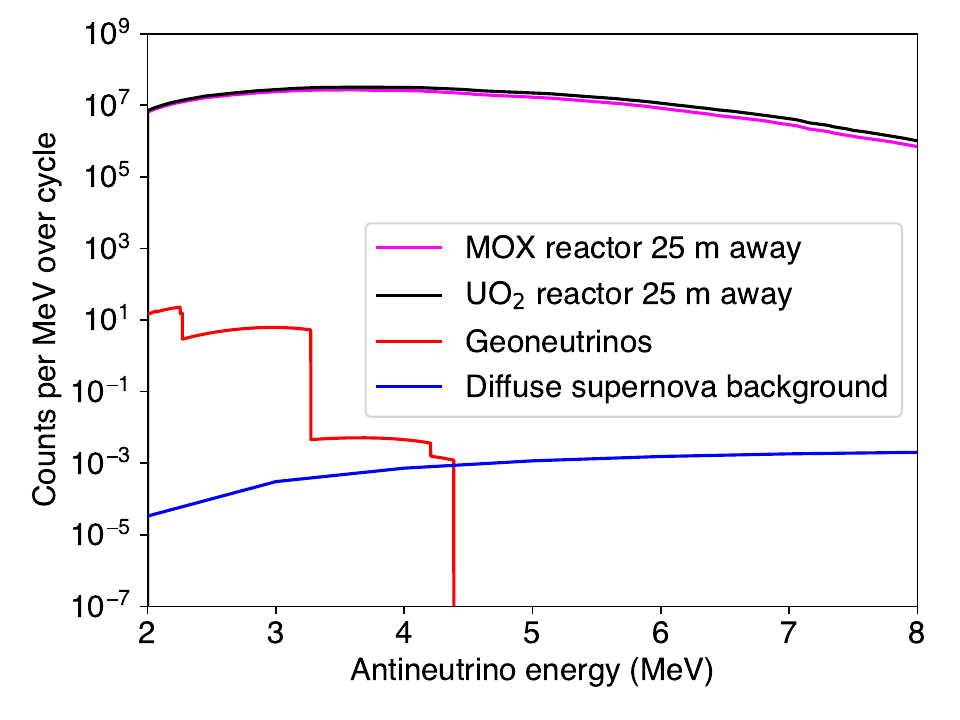}
    \caption{Background antineutrino counts (red, blue) over 50~GWD/tHMi irradiation time in the 2--8~MeV energy range using data from Ref.~\cite{Vitagliano_2020} summed over flavors/masses, compared to the reactor counts (black, magenta).}%
    \label{fig:background}%
\end{figure}

Calculated using the procedure discussed in Section~\ref{sec:detectionmodeling}, Figure~\ref{fig:background} shows the energy-dependent detector counts resulting from these sources integrated over time for a 25~m reactor-detector distance using detector characteristics introduced in Section~\ref{sec:detectionmodeling}. We also show in this plot the energy spectrum and count rate from reactor neutrinos of our baseline setup. 
The number of detected reactor neutrino events exceeds the number of detected background neutrinos at the chosen distance of 25~m by a factor of $\sim$10$^7$.
In fact, the background count is less than 1\% of the reactor count up to a distance of 20~km.
Therefore, in the case where the detector is 25~m from the core, it is safe to ignore antineutrino background sources.

\noindent
{\it Non-neutrino induced background}

Although we may neglect non-reactor neutrinos as possible background, cosmogenic sources copiously produce confounding correlated IBD-like detector events.
Cosmic rays include muons and hadrons, especially fast neutrons, which can produce detector signals closely resembling antineutrinos.
All of these background sources can, in principle, be discerned from true antineutrino events with the degree of discrimination that depends on the event selection criteria (which typically impacts the detection efficiency) and by passive shielding around the detector.

Neutron scattering on hydrogen followed by neutron capture results in two separate events that can mimic those from IBD, with some differences in the energy, time profiles, and topology.
We can use these differences to distinguish true antineutrino events as best we can, but it is better to simply allow fewer hadrons to reach the detector.
Many neutrino detectors are built underground for this reason, and additional plastic (``low-Z'' for low proton number) and heavy metal shielding can be added around the scintillator.

Cosmic ray muons may be captured by nuclei, either inside the detector or in the surroundings, resulting in secondary fast neutrons or pions by $\mu N\to \mu N'+X$ inelastic scattering.
They may also decay, resulting in another, real electron-flavored antineutrino (that is unlikely to be in the same energy range as from the reactor) and an electron that loses energy into photons similar to the positrons from IBD.
Strategic scintillator design, configuration, and data analysis tracks muon paths alongside their scattered hadrons within the detector, allowing for their identification and dismissal from the pool of possible antineutrino-related events;
this is often referred to as a muon veto system.

Hadrons resulting from spallation that occurs outside the shielding can be attenuated by the heavy metal surrounding the detector, just like for the primary neutron and pion cosmic rays.
However, these same shields are rife with nuclei for muons to interact with, producing hadrons that have less shielding to prevent them from reaching the detector.
Therefore, neither a muon veto system nor the shields can perfectly prevent their associated background, and there is an optimum amount of shielding that maximizes the background rejection reliability of each.

The simplest way to determine the background rate is to operate the detector in the same conditions (all shielding present, correct location on reactor site, \textit{etc.})\ while the reactor is off.
It is also possible to predict non-antineutrino-induced background by using the Monte Carlo simulations of particle interactions in matter, \textit{e.g.,} \textsc{Fluka}~\cite{FLUKA} and \textsc{Geant}4~\cite{GEANT4}.
The results will depend on the structure and materials of a particular detector.

In the current study, total background counts are included as one of the independent variables that determine the probability of discerning a MOX discrepancy.
The background count rate can be as high as an order of magnitude above the maximum possible reactor neutrino count rate without selection and shielding, and as low as half of the count rate when we implement the above discernment techniques~\cite{Alekseev_2014}.
Ultimately, the MOX discrepancy determination ability depends on the amount of possible background reduction and on the efficiency of detecting true antineutrino events.

\subsection{Neutrino Detection Count Modeling}
\label{sec:detectionmodeling}
We model neutrino detection based on a segmented plastic scintillator design similar to the DANSS~\cite{Alekseev_2014} and PANDA~\cite{Kuroda_2012} detectors. 
These detectors are designed to use approximately one ton of scintillator active volume.
However, the effectiveness of the methods employed in this study is limited by uncertainties due to counting statistics, and thus larger detectors have more sensitivity to provide clear signals of small changes resulting from the diversion of a fraction of the fuel. 
Therefore, we assume a detector with five tons of plastic scintillator with the characteristics of the EJ-260~\cite{EJ260}---a commercially available plastic scintillator similar to those used in several neutrino detectors, \textit{e.g.}, Ref.~\cite{haghighat2020observation, Dorrill_2019,Netrakanti_2022}.
The IAEA has stated that any neutrino detector designed for nonproliferation should be capable of transportation in an ISO container~\cite{IAEA_final_report_2008}.
The volume of this container would be able to fit a five-ton plastic scintillator detector along with the associated electronics and shielding.

The detected neutrino count collected over time can be modeled by \begin{equation}\label{eqn:N_nu}
        N_\nu = \frac{N_p}{4\pi D^2} \sum_\ell \int_0^T \text{d} t\ N_\ell\left(t\right) \int_{1.806 \text{ MeV}}^{8 \text{ MeV}} \text{d} E\ \sigma\left(E\right) \phi_\ell\left(E\right) \epsilon\left(E\right).
    \end{equation}
Here, $N_\ell$ is the time-dependent isotope-specific fission rate of nuclide $\ell$ in the hypothetical reactors described in Sect.~\ref{sec:nmodel}.
Introduced in Sect.~\ref{sec:nspec}, $\phi_\ell(E)$ is the neutrino emission spectra per fission for nuclide $\ell$.
Next, $\sigma(E)$ is the IBD cross-section~\cite{Vogel_Beacom_1999} for an electron antineutrino on proton interaction, which is steeply rises with incident antineutrino energy.
The detector efficiency $\epsilon$ represents the fraction of IBD events that result in a detected count.
The number of free protons in the detector $N_p$ is determined by the scintillating material, in this case, the EJ-260~\cite{EJ260}.
The distance between the reactor and the detector $D$ is taken to be 25 meters, the approximate distance from the core to the tendon gallery.
The energy integral spans from 1.806~MeV, the IBD threshold, to 8~MeV, the maximum energy of the available reactor antineutrino spectral data, and the time integral is applied over the measurement period $T$.
The detector efficiency is in general a function of energy, but in this study, we treat it as a constant for simplicity.

\begin{figure}[t!]
    \centering
    \includegraphics[scale = 0.5]{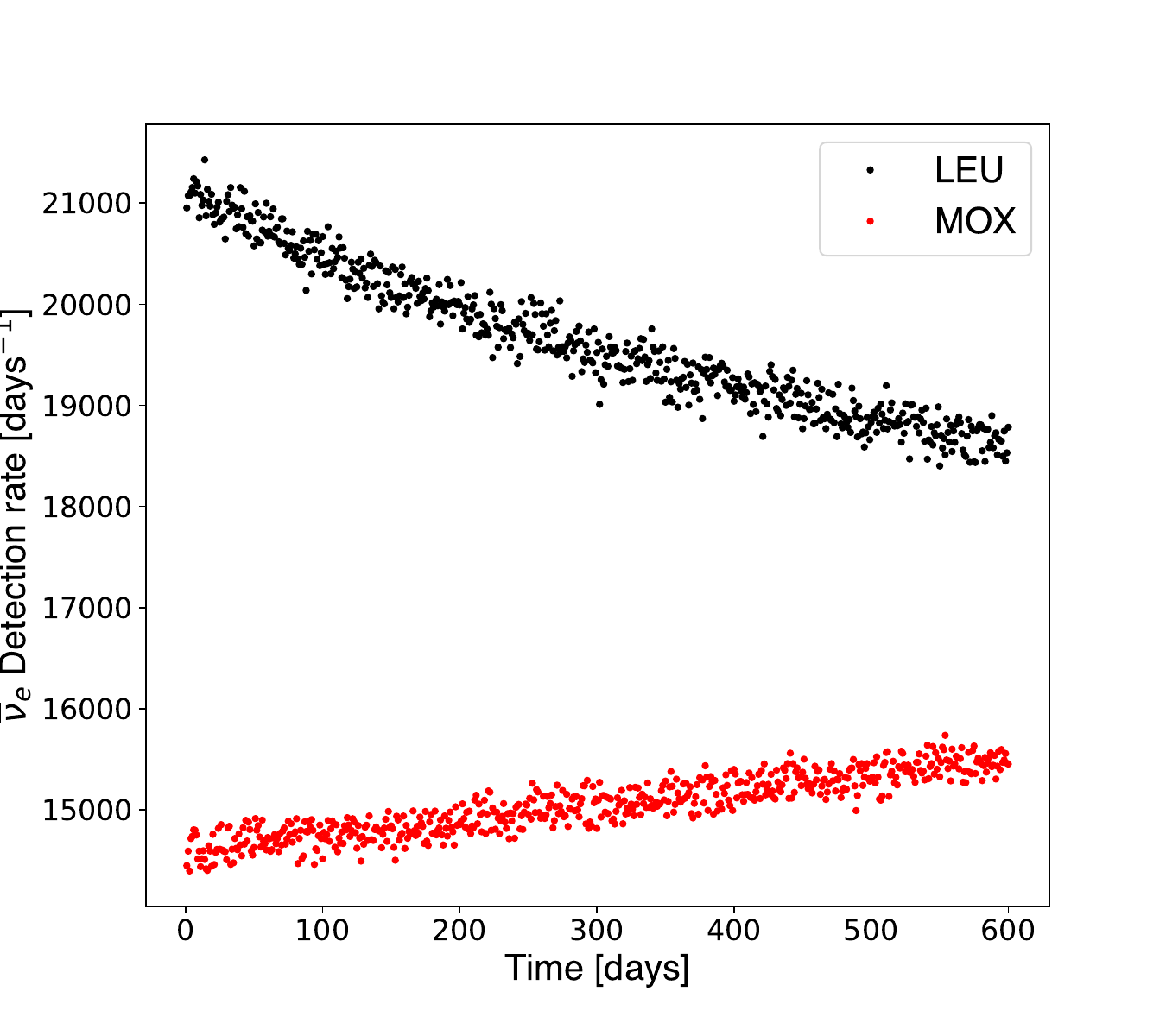}
    \caption{Two trials and the resulting the number of neutrinos detected at 100\% detection efficiency. The trial labeled LEU results from a core containing entirely low-enriched uranium, while the trial labeled MOX results from a core containing entirely MOX fuel.}
    \label{fig:sample_neutrino_observation}
\end{figure}

To create mock trial data for a reactor with a certain MPF in one-day interval time steps, the neutrino counts are sampled using the result of the Eq.~(\ref{eqn:N_nu}) as the mean at each time step and assuming a Poissonian variance. 
The simulated background count rates are also sampled from a normal distribution with a Poissonian variance.
We approximate the mean background rate as a fraction of the average daily neutrino detection rate for a full LEU core.
We vary this fraction and explore its effect on the statistical test.
Systematic uncertainties related to the neutrino spectrum model are included but are subdominant compared to statistical fluctuations.
\Cref{fig:sample_neutrino_observation} shows two examples of sampled mock data using 100\% detection efficiency.

\section{Detecting Anomalous Activity}
\subsection{Test Statistic Formulation}
\label{sec:tsf}

\begin{figure}[t!]
    \centering
    \includegraphics[scale = 0.4]{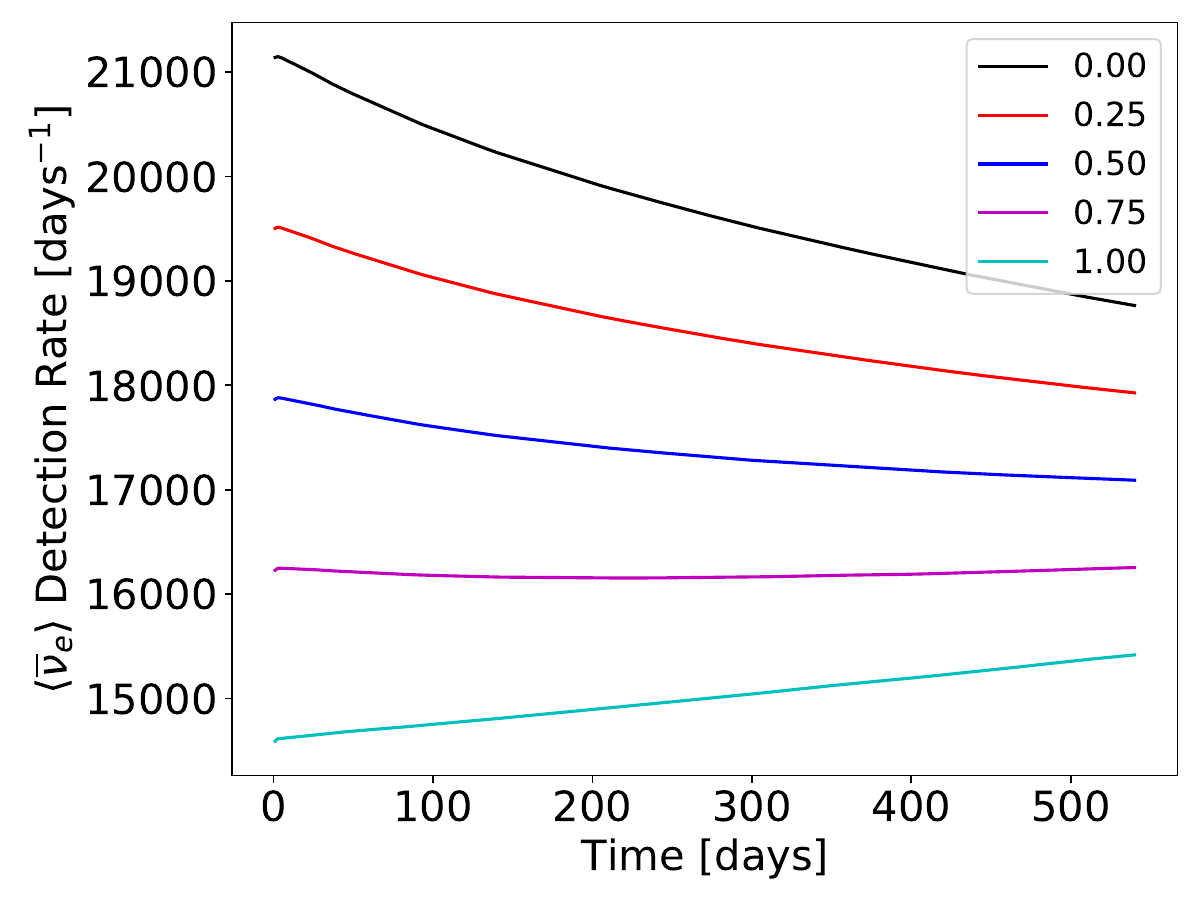} 
    \includegraphics[scale = 0.4]{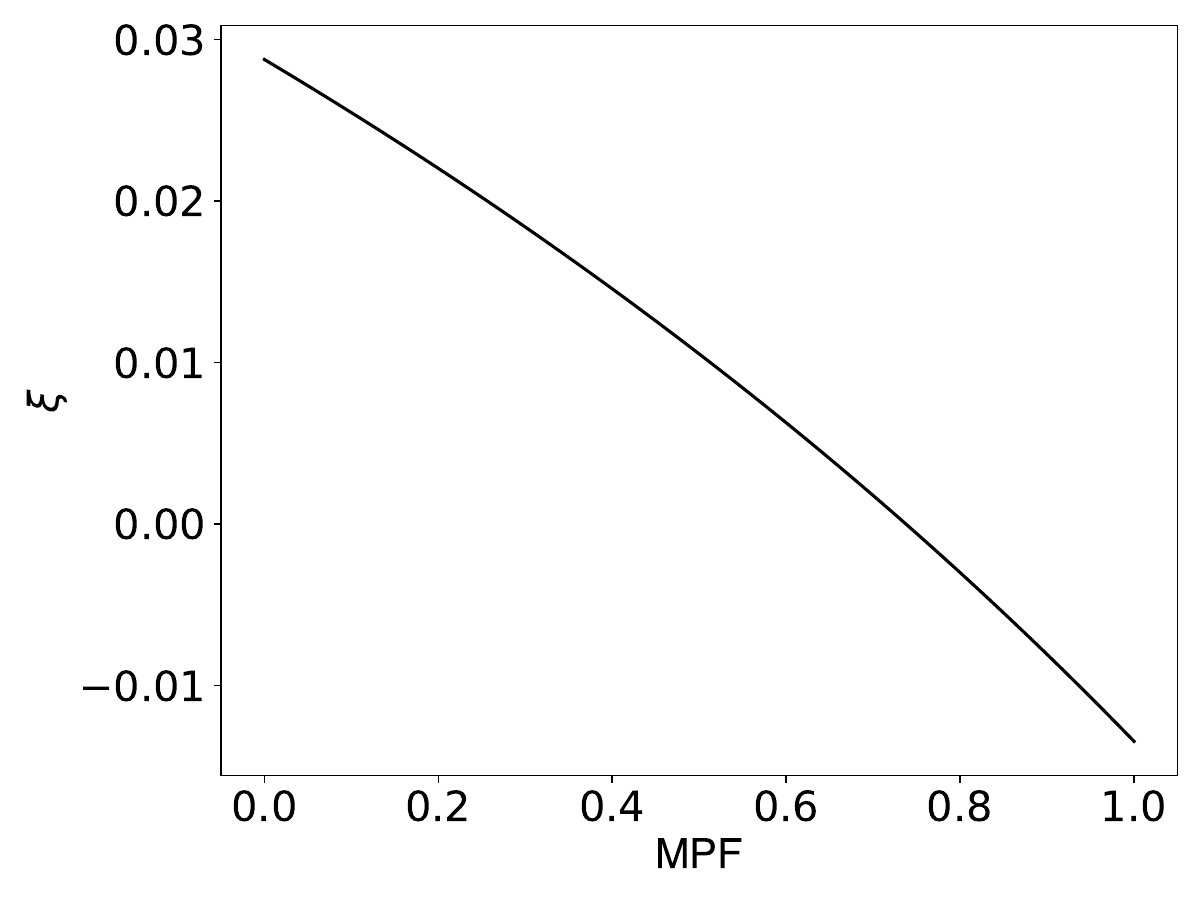}
    \caption{Left: Expected neutrino count rate evolution for various MPFs. A 100\% efficient detector is assumed. Right: Expected value of $\xi$ as a function of MPF, assuming 100\% detector efficiency, and zero background over a 540-day observation period. 
    }
    \label{fig:emissions_vs_time}
\end{figure}

We propose a statistical test in which we compare a quantity $\xi$ between a declared and simulated MOX loading fraction and determine the probability that the measured (simulated) data matches the declared (predicted) expectation.
This represents a realistic application of the test in which a reactor operator declares the characteristics of their reactor, and an inspector compares the measured neutrino counts with what is predicted from the declarations and determines the probability of anomalous activity.
We define the test statistic $\xi$ as
\begin{equation}
    \label{eq:xi}
    \xi = \frac{n_1 - n_2}{n_1 + n_2 - \overline{Y}},
\end{equation}
where a count measurement is taken over a period $\Delta t$ {from the beginning of the reactor cycle}; $n_1$ is the neutrino count over the first half of this period, $n_2$ is the number of neutrino counts over the second half, and $\overline{Y}$ is the expectation of the number of background counts during $\Delta t$.
This background rate could be measured while the reactor is offline.
This quantity $\xi$ accounts for the total counts and fuel evolution of the entire measurement period, which depends on the power of the reactor and the fuel being used.
The left side of Fig.~\ref{fig:emissions_vs_time} shows the time evolutions of detected neutrino rate for various MPFs, which can be explained by considering the fission rates in Fig.~\ref{fig:relfiss} and known neutrino outputs for specific fission processes.
The quantity $\xi$ trends monotonically with the MPF, shown on the right side of Fig.~\ref{fig:emissions_vs_time}, making it a useful measurement of the amount of MOX fuel in the reactor.
The statistical significance of the difference between the measured and predicted value can be quantified by the uncertainty in the predicted value, and a $p$-value can be determined, indicating the probability of compatibility between the two values.

\subsection{Test Statistic Uncertainty Propagation}
\label{sec:tsup}
There are several sources of uncertainty to account for.
The statistical uncertainty is straightforwardly Poissonian because of the random timing of the fissions that produce the neutrinos.
The most significant systematic uncertainties originate from the neutronics simulation and the neutrino spectral models.
Aside from the negligible time-dependent isotope-specific uncertainties in fission reaction rates reported by the Monte Carlo neutron transport calculation with the Serpent 2 neutronics code, there are also cross section uncertainties in the nuclear libraries used for the simulation.
Reference~\cite{osti} propagates the cross-section covariance data through burnup to determine the effect on isotopic composition at a final burnup of 50~GWD/MTHM.
The coefficient of variation describing the uncertainty distribution in isotopic composition is the same as the uncertainty distribution in the fission rate.
To construct a burnup-dependent uncertainty we assume the cross-section uncertainty starts at zero at 0~GWD/MTHM and increases linearly to match the 50~GWD/MTHM uncertainty reported in Ref.~\cite{osti}.
The total uncertainty of the fission rates is propagated as a source of systematic error through the neutrino count model in Eq.~(\ref{eqn:N_nu}).

The PRL2023 neutrino spectral model's discrepancy from the true neutrino spectrum is quantified by uncorrelated energy-dependent, isotope-specific uncertainties.
We take a Monte Carlo approach to determine the resulting systematic variance in detected neutrino counts.
After separately determining all of these time-dependent contributions to the variance of the neutrino detection rate, they can be added in quadrature to find the total uncertainty in $n_1$ and $n_2$, which we then propagate to determine the uncertainty of $\xi$.

\subsection{Detection of Anomalous Fuel Loading}
This test is designed to detect anomalous fuel loading in nuclear reactors partially loaded with MOX fuel. 
The enrichment process and production of weapons-grade plutonium are not the primary concern of this study.
Instead, we are concerned with the diversion from a reactor of fresh MOX fuel with a plutonium content that would be considered weapons-grade.
We model this diversion scenario using the method described in Sect.~\ref{sec:nmodel}.
Diversion results in a discrepancy in neutrino count rate evolution from our expected counts, which our statistical test is capable of observing.
We quantify this method's resolving power in Sect.~\ref{sec:discpower}.

\begin{figure}[t!]
    \centering
    \includegraphics[scale = 0.5]{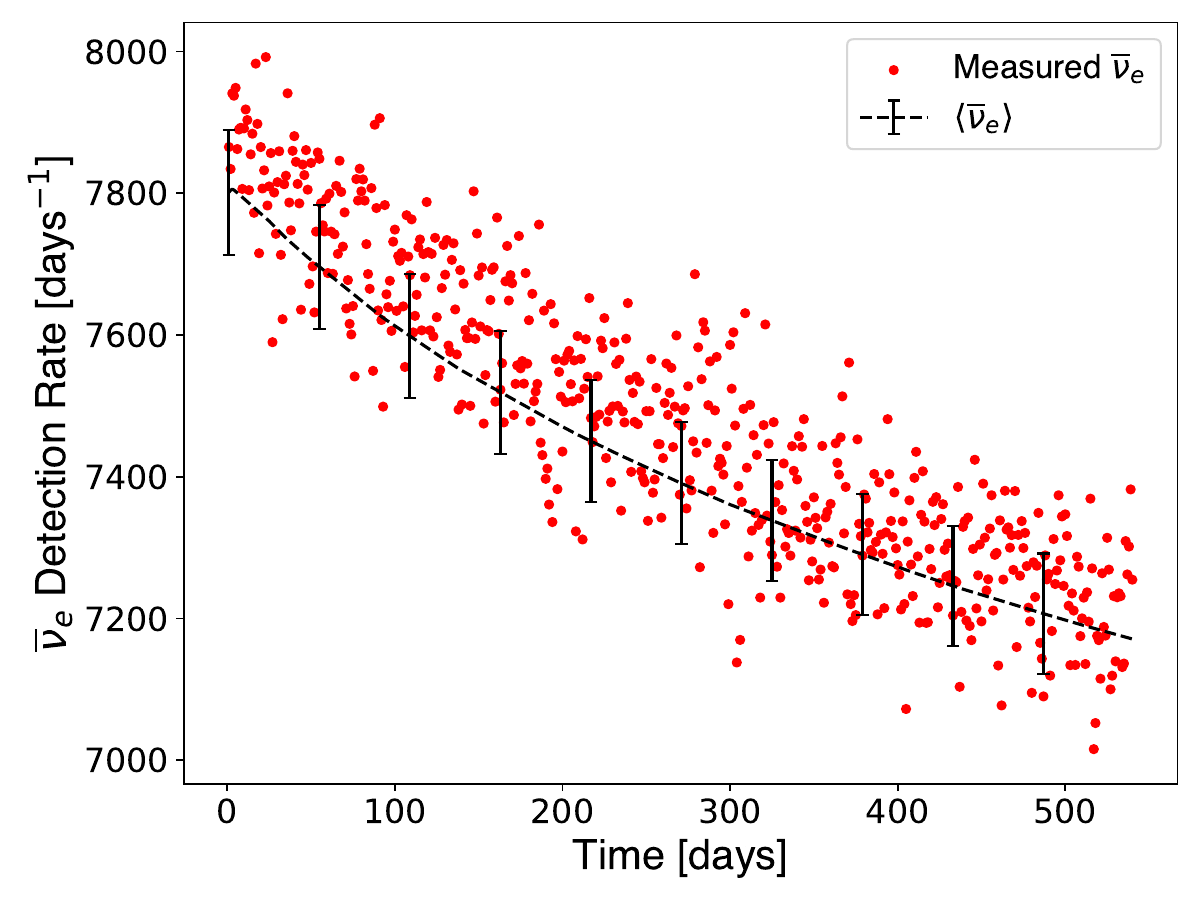}
    \caption{Simulated measurement of the neutrino count rate for a reactor with an MPF of 0.22 for a 540-day observation period corresponding to the true core (red), and the expected count rate and associated uncertainty for a reactor with an MPF of 0.25 corresponding to the declared core (black). This scenario corresponds to a neutrino detector with a 40\% detector efficiency.}
    \label{fig:example}
\end{figure}

As a demonstration of how this statistical test functions, consider the following case.
A reactor facility declares the enrichment, quantity, and positional distribution of fuel within a partially loaded MOX core. 
These declarations are simulated using a high-fidelity nuclear transport code, and the equivalent MPF for these declarations is found to be 0.25, meaning 25\% of the total reactor power comes from MOX-loaded fuel pins.
The isotope-specific time-dependent fission fractions determined by the simulation are used in Eq.~(\ref{eqn:N_nu}), along with the HM spectral model and detector material and size, to determine the expected neutrino detection rate.
With this rate, the expected value of $\xi_\text{declared}$ is determined for a total observation period of 540 days, the typical operation time of a reactor between refueling.
Meanwhile, suppose the reactor operator chooses to divert 12\% of the MOX fuel to a nuclear weapons program, corresponding to about 120~kg of WGPu.
In an attempt to preserve cycle lengths, the MOX fuel is replaced with LEU fuel.
With this new fuel configuration, the MOX power fraction has changed to an MPF of 0.22.
The reactor operates under these conditions for the duration of the measurement period.
\Cref{fig:example} depicts the a sample run of the number of neutrinos measured each day during the observation period compared to the declaration-based count rate expectation.
%

For each of these trial sets of data, $\xi$ is calculated and compared using significance variable $z$:
\begin{equation}
    \label{eqn:compare_xi}
    z = \frac{\xi_\text{measured} - \xi_{\text{declared}}}{\sigma_{\xi_\text{declared}}}.
\end{equation}
In our example case, \cref{eqn:compare_xi} gives $z=2.45$, corresponding to a $p$-value of $\approx 0.014$;
we can determine that the difference between reactor operations and declarations is statistically significant. 
In this study, we simulate reactors with measured MPFs lower than the declared to determine our sensitivity to the diversion of fresh MOX fuel, corresponding to an illegitimate removal of weapons-grade plutonium.

\section{Results}

\subsection{Quantification of Discriminatory Power}
\label{sec:discpower}
\begin{figure}[t!]
    \centering
    \includegraphics[width = .65\textwidth]{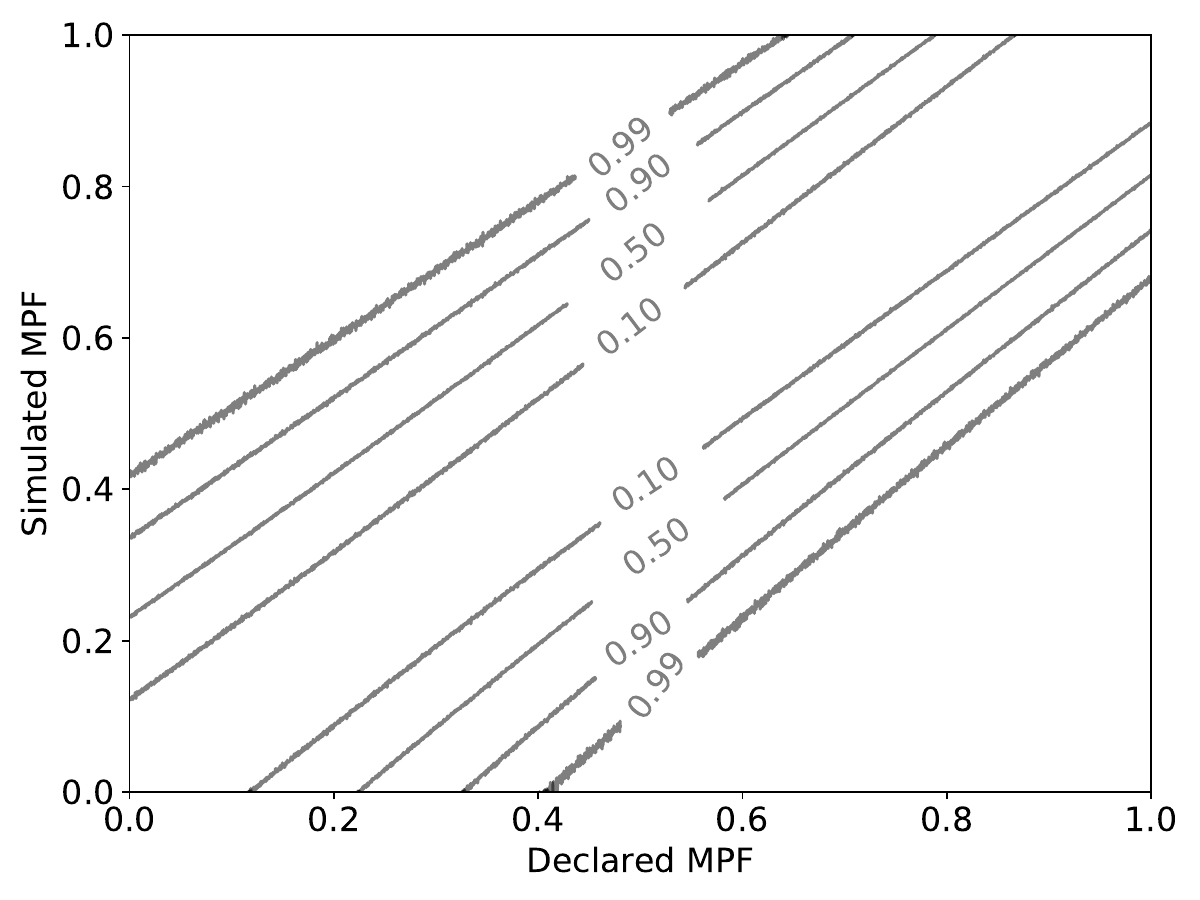}
    \caption{Fractions of $10^4$ trials that successfully distinguish declared and true MPF at 95\% confidence using a detector with 20\% efficiency.
    Here we assume 30\%  background to calculate the expected neutrino count.}
    \label{fig:contour}
\end{figure}

\begin{figure}[t]
    \centering
    \includegraphics[scale = 0.4]{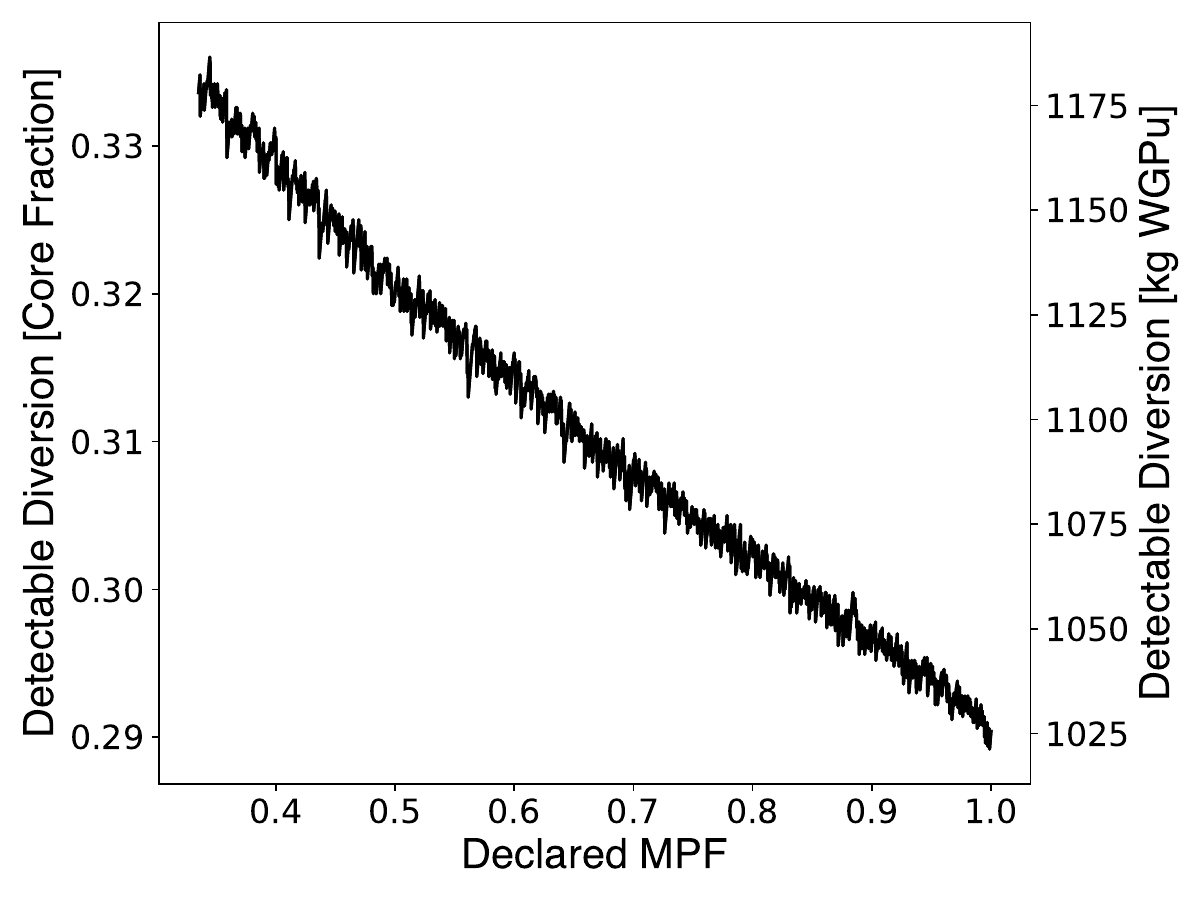}\includegraphics[scale = 0.4]{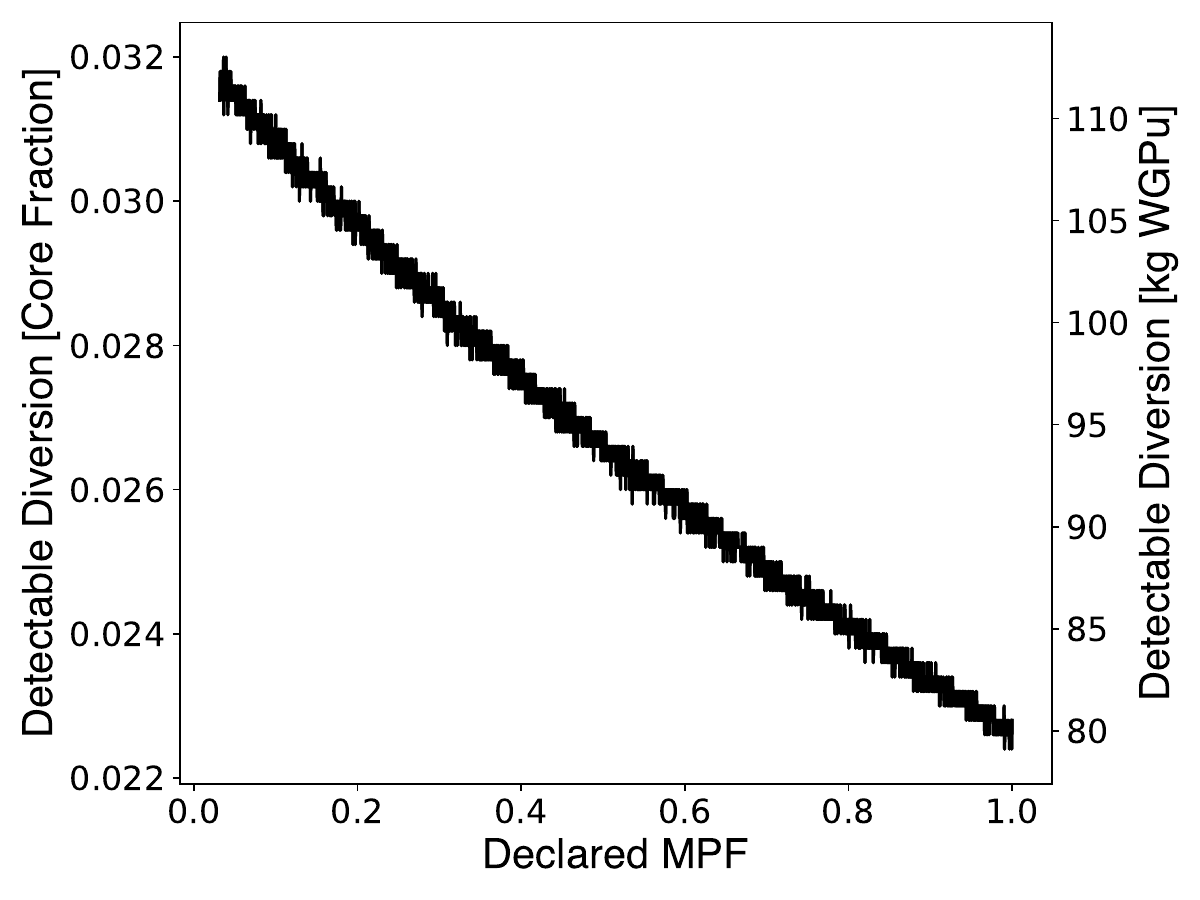}
    \caption{MPF resolution, the minimum divertible core fraction and plutonium mass that can be detected in 90\% of trials. Left: detector with 20\% efficiency, 30\% background, and for a relatively short measurement period of 180 days. Right: detector with 100\% efficiency, no background and a measurement period of 18 months (540 days).}
    \label{fig:lineplot}
\end{figure}

We can visualize the resolving power of the statistical test
using a contour plot depicted in Fig.~\ref{fig:contour}, with contours labeled by the fraction $f$ of tests with statistically significant results.
The distance from a contour to the identity line is the MPF ``resolution,'' the difference between declared and simulated MPF that is distinguishable at the 95\% confidence level in a fraction $f$ of total trials, shown for $f=90\%$ in Fig.~\ref{fig:lineplot}. 
Since we are primarily concerned with the diversion of fresh MOX fuel, this is specifically the distance between the \textit{lower} 90\% contour  in Fig.~\ref{fig:contour} and the identity line (the contours are not symmetric across it).
By assuming that each fuel rod in the reactor is producing equal power, this core fraction is related to the diverted mass of weapons-grade plutonium by a proportionality constant that we determine by using fuel rod characteristics from \cite{kozlowski2006pwr}.
Our discriminatory power then is quantified by the inverse of the resolution for a given pair of MPFs, detector, and observation period, i.e.\ if we can detect a smaller diversion, our discriminatory power is greater.

\subsection{Effect of Detector Efficiency}
\begin{figure}[t!]
    \centering
\includegraphics[scale = 0.4]{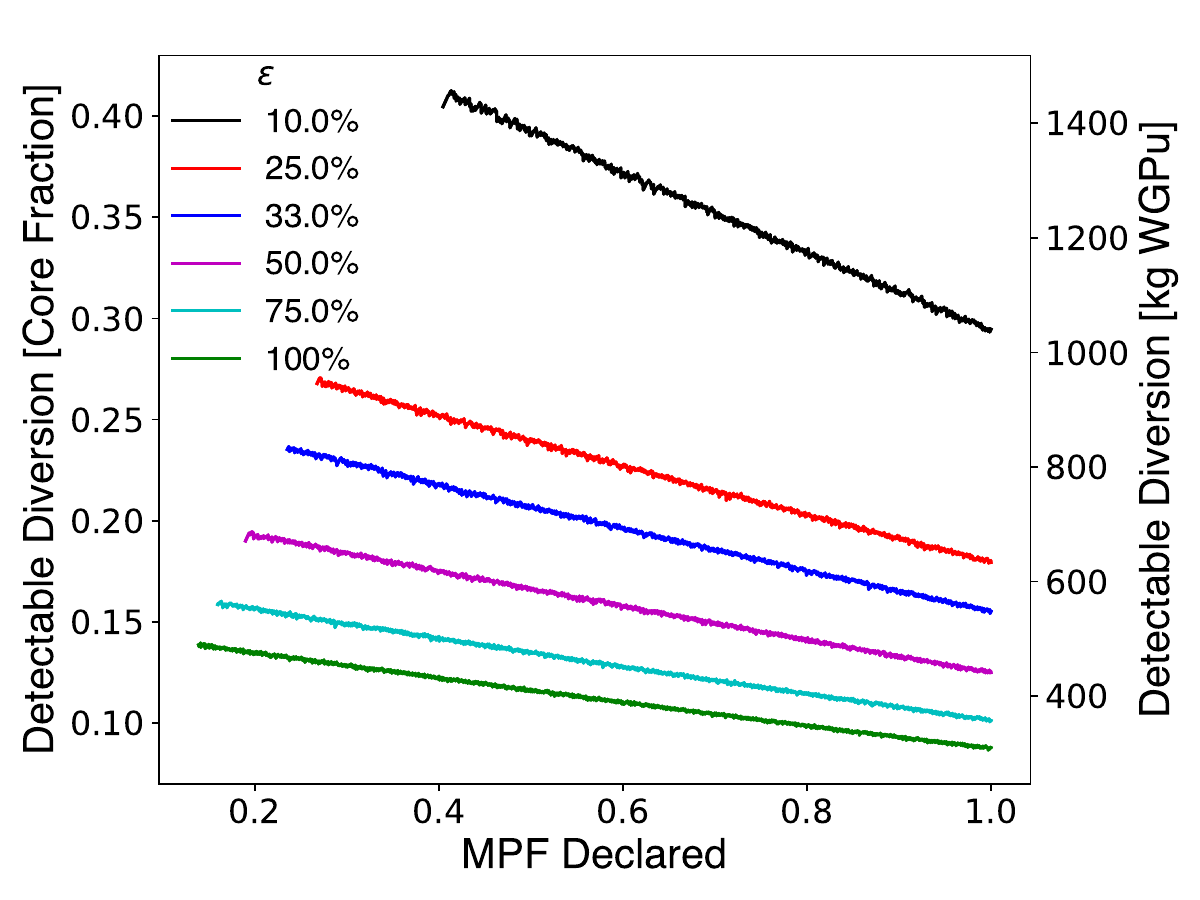}
    \caption{Minimum divertible core fraction and plutonium mass that can be detected in 90\% of $10^4$ trials for various detector efficiencies.
    Each test uses an observation period of 180 days and no background.}
    \label{fig:lineplot_efficiency}
\end{figure}

\begin{figure}[t!]
    \centering
\includegraphics[scale = 0.4]{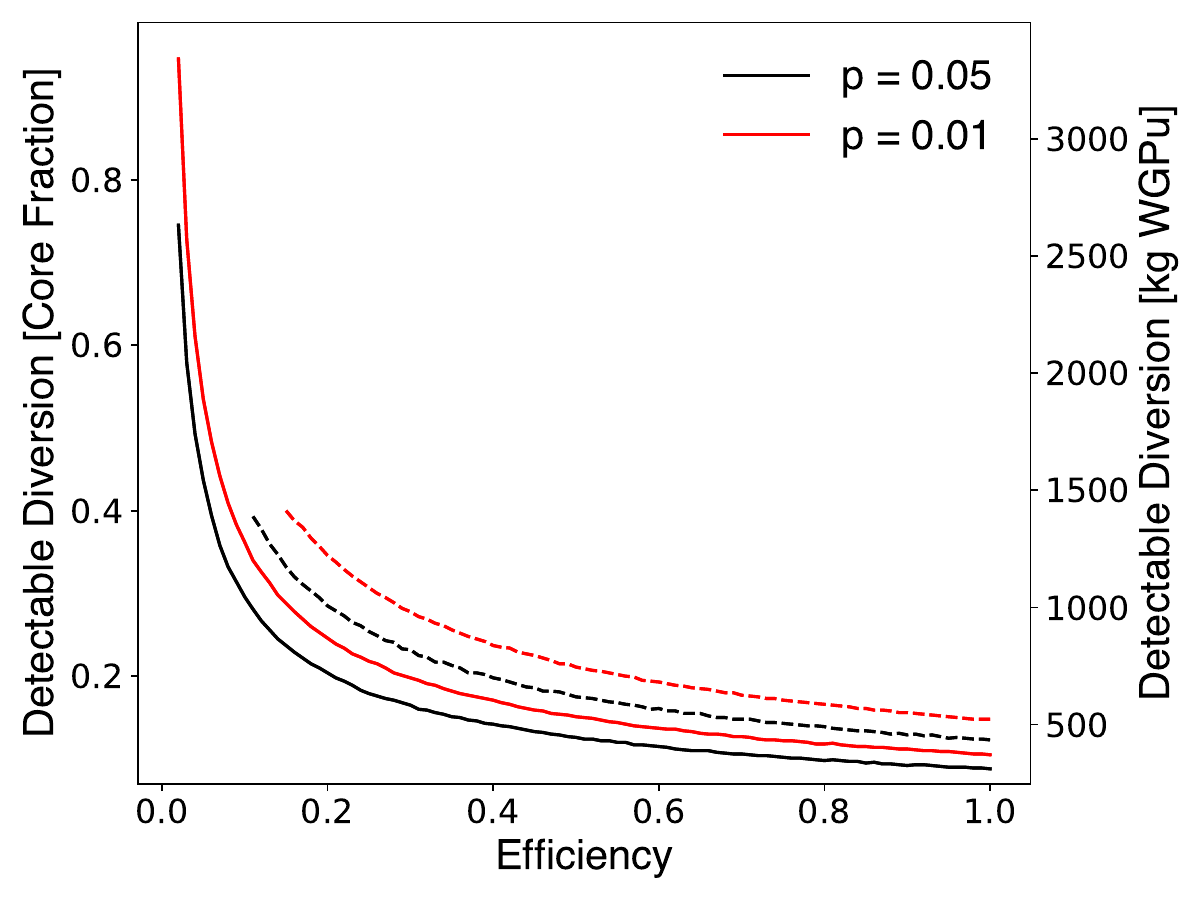}
    \caption{Minimum divertible core fraction and plutonium mass that can be detected in 90\% of $10^4$ trials depending on detector efficiency. The declared MPFs are 0.4 (dashed) and 1 (solid). }
    \label{fig:resolution_vs_efficiency}
\end{figure}

We would like to explore the sensitivity of our statistical test to detector efficiency, which ranges from about 5\% to near 50\% in modern neutrino detectors \cite{Netrakanti_2022, SoLid}.
We calculate diversion resolution for several different detector efficiencies in Fig.~\ref{fig:lineplot_efficiency}.
For each line shown, the underlying test assumes a zero background rate and a 180-day measurement period. 
As one might expect, a lower efficiency leads to a less sensitive measurement because of fewer detected events. 
Figure~\ref{fig:resolution_vs_efficiency} shows the resolution vs efficiency for two different MPFs.
As expected, the discriminatory power is proportional to the number of events observed.

\subsection{Effect of Background}
Next, we explore the effect of different background rate on diversion resolution. 
We treat the background as a rate sampled on each day from a Gaussian distribution with a coefficient of variation of unity.
We assume a detector with a 20\% efficiency and use an observation period of 180 days. 
We quantify the background rate as percentage of the average neutrino detection rate over the observation period for a reactor with an MPF of 0, or a total LEU core, corresponding to signal-to-background ratios ranging from 0.2 to 1~\cite{Bulaevkaya_Bernstein}.

\begin{figure}[t!]
    \centering
\includegraphics[scale = 0.4]{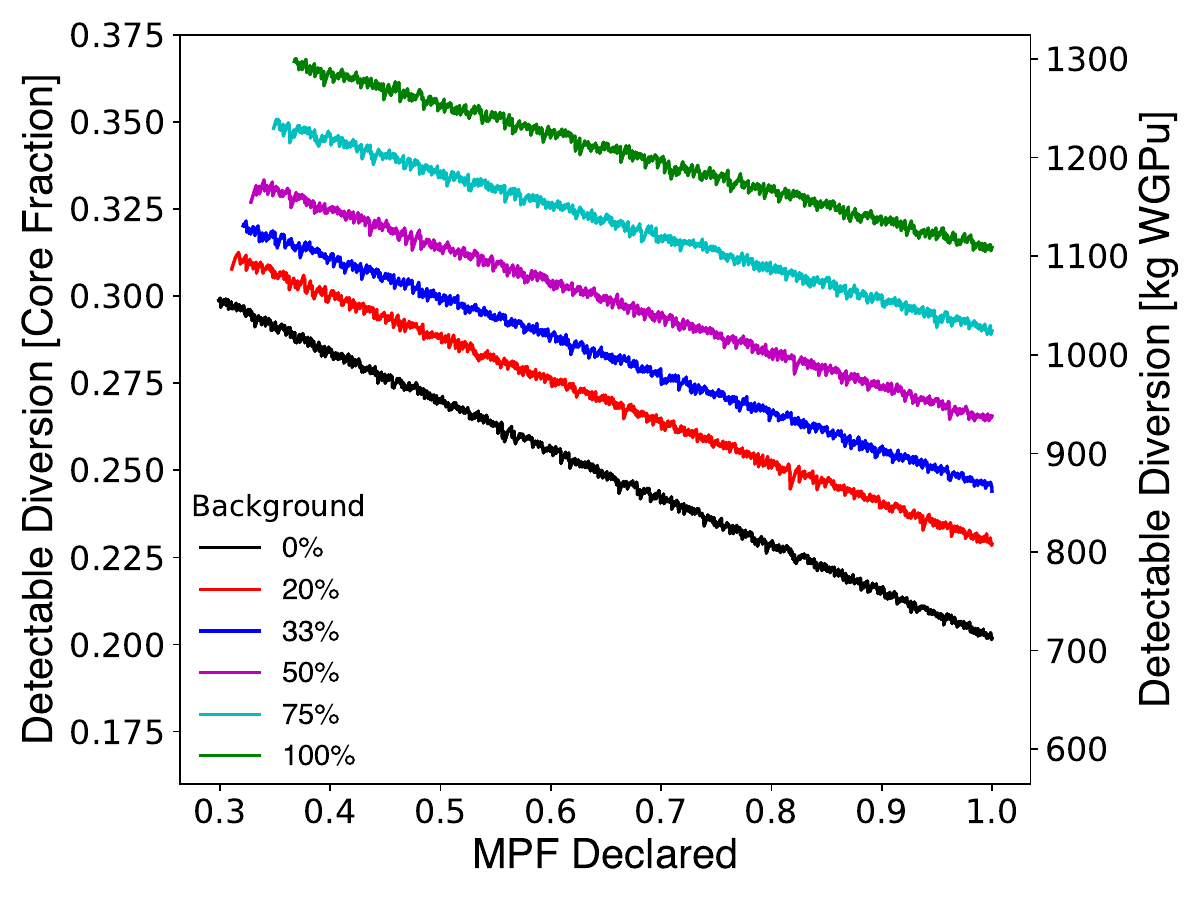}
    \caption{Minimum divertible core fraction and plutonium mass that can be detected in 90\% of $10^4$ trials for various background rates. Each test uses a detector efficiency of 20\% over an observation period of 180 days.}
    \label{fig:lineplot_bg}
\end{figure}

\begin{figure}[t!]
    \centering
\includegraphics[scale = 0.4]{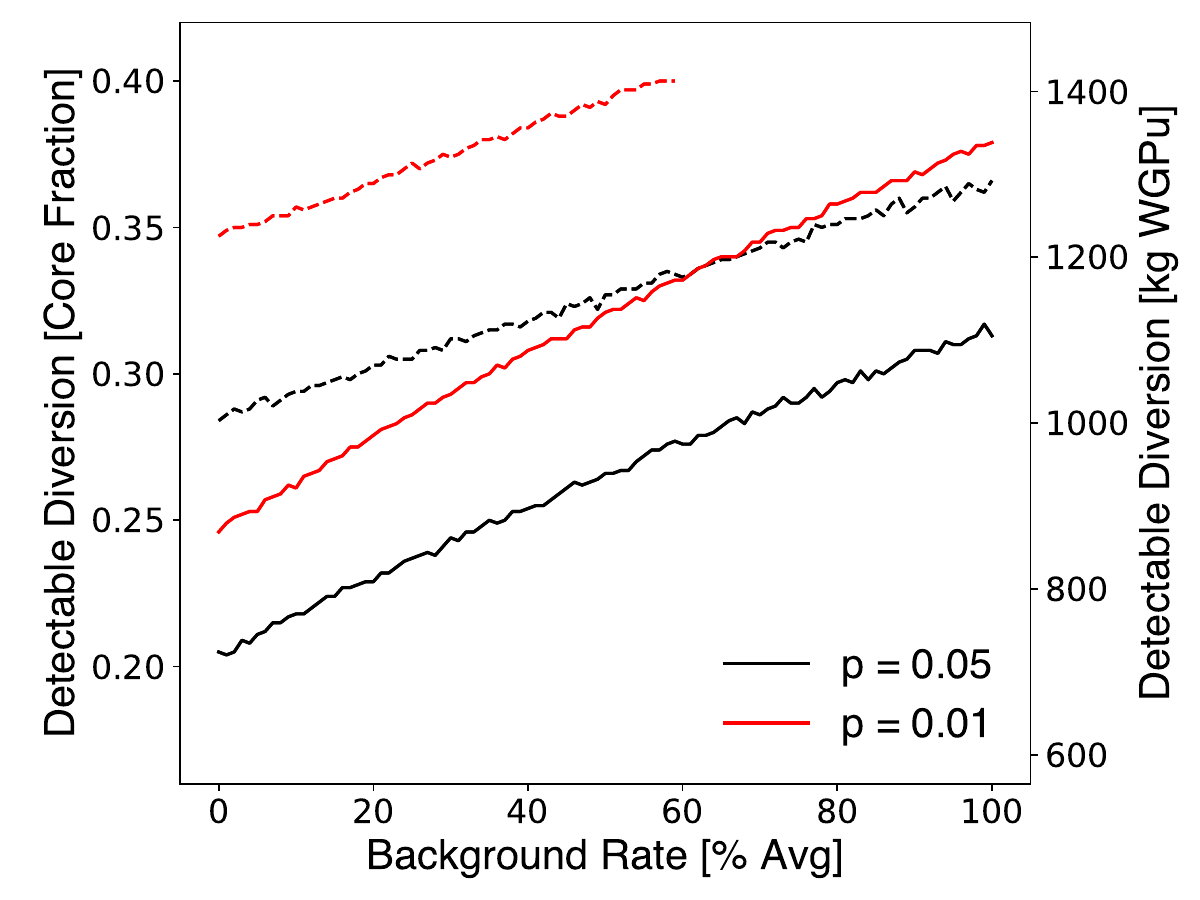}
    \caption{Minimum divertible core fraction and plutonium mass that can be detected in 90\% of $10^5$ trials depending on background rate. The declared MPFs are 0.4 (dashed) and 1 (solid). }
    \label{fig:resolution_vs_bg}
\end{figure}

\Cref{fig:lineplot_bg} shows the line plot of diversion resolution as a function of MPF for several background rates.
Naturally, a higher background rate leads to a worse discriminatory power because of the increased statistical uncertainty and a smaller change in the count rate with observation time.
\Cref{fig:resolution_vs_bg} depicts minimum detectable diversion versus the background rate.

\subsection{Application to Continuous Monitoring}

Next, we explore the effect of different observation periods.
We use a detector with a 20\% detector efficiency and no background. 
We explore observation periods ranging from 45 days to 540 days;
five hundred and forty days is the average amount of time between reactor refuelings.

\begin{figure}[t!]
    \centering
\includegraphics[scale = 0.4]{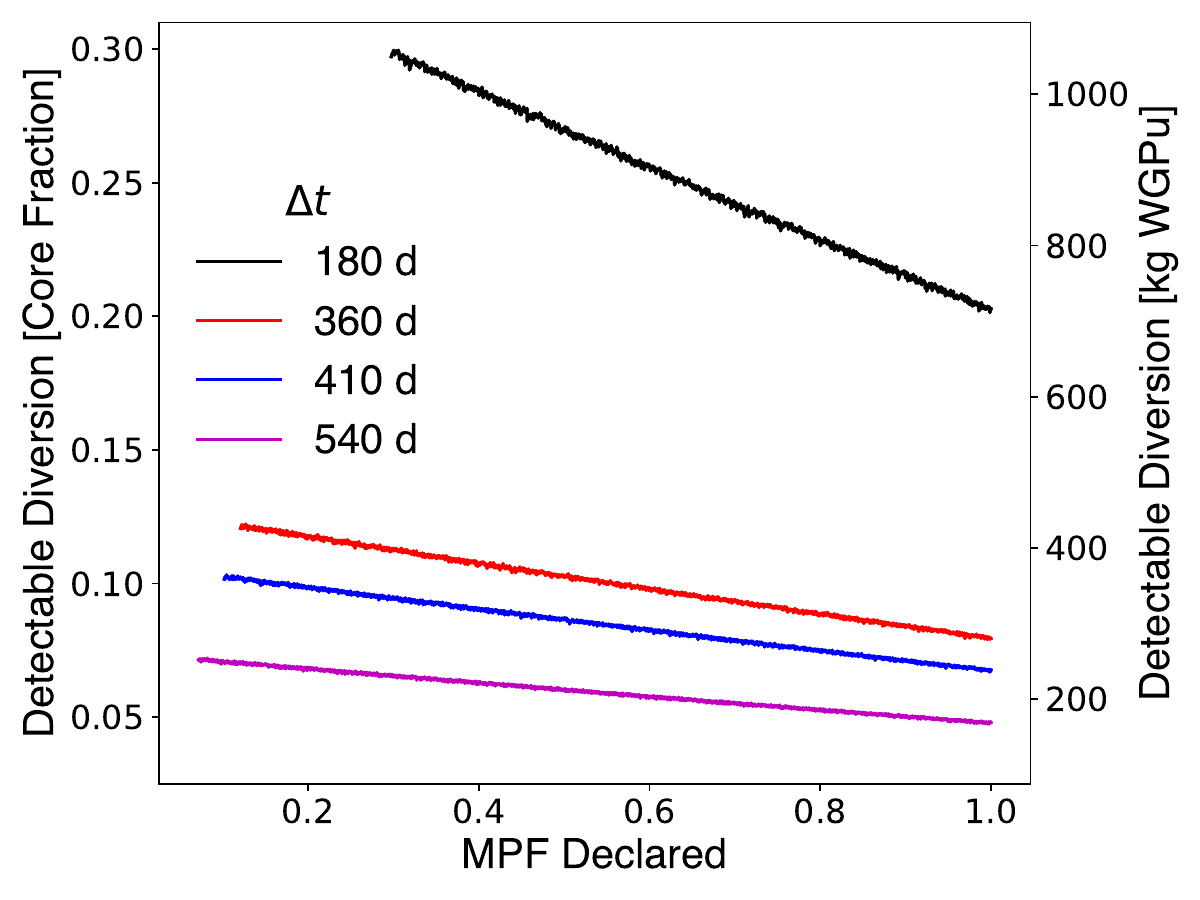}
    \caption{Minimum divertible core fraction and plutonium mass that can be detected in 90\% of  $10^4$ trials. Each test uses a detector efficiency of 20\% and no background. }
    \label{fig:lineplot_dt}
\end{figure}

With an observation period of 45 days, more than half of the core could be diverted without our test distinguishing it. \Cref{fig:lineplot_dt} shows the results for several different longer observation periods, where the discriminatory power improves. For a measurement period of 540 days, the test can distinguish between reactors with a 7\% difference in fuel loading. 

\begin{figure}[t]
    \centering
\includegraphics[scale = 0.4]{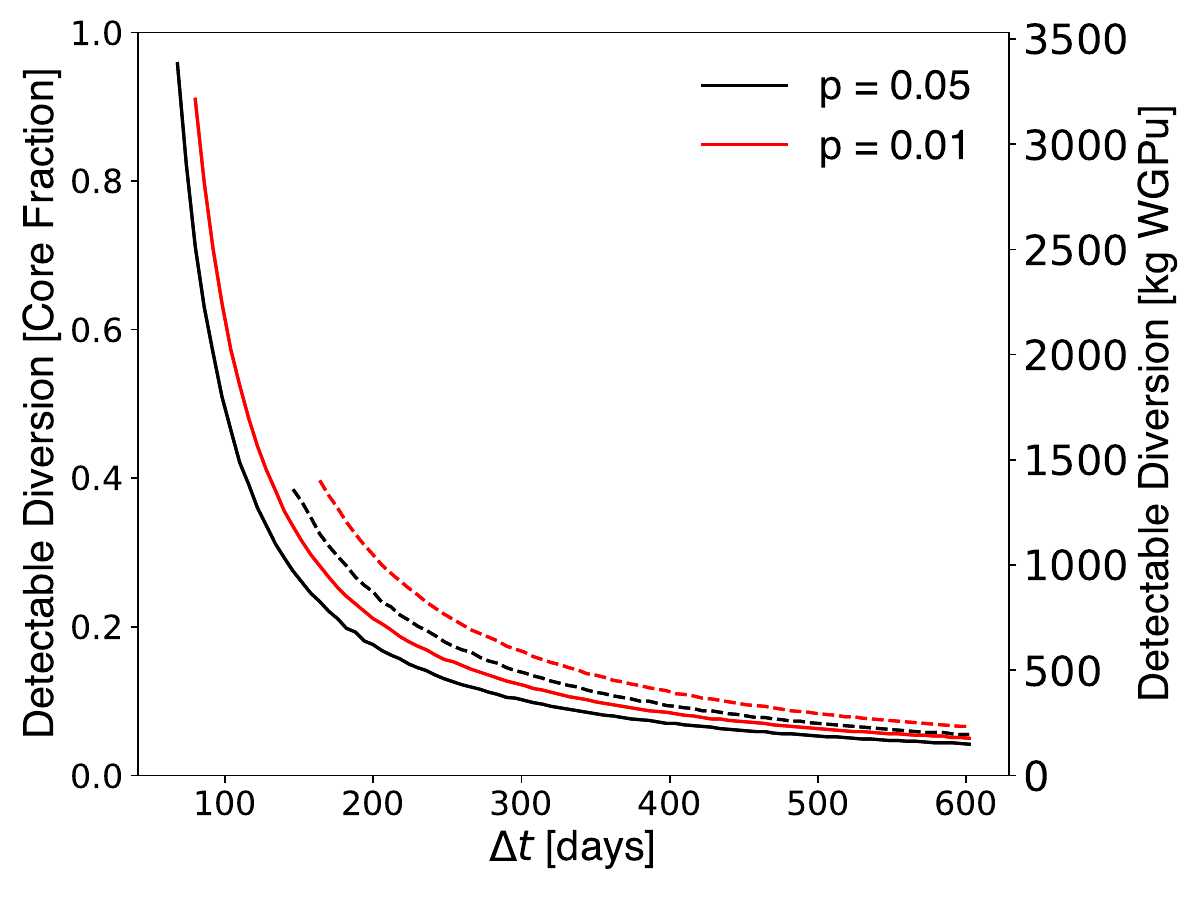}
    \caption{Minimum divertible core fraction and plutonium mass that can be detected in 90\% of $10^4$ trials depending on observation period.  The declared MPFs are 0.4 (dashed) and 1 (solid). }
    \label{fig:resolution_vs_dt}
\end{figure}

\Cref{fig:resolution_vs_dt} shows the relationship between the maximum resolution vs MPF and the measurement period. 
We explored 100 different values of $\Delta t$ ranging from 6 days to 600 days.
The divertable core fraction is indeed inversely proportional to the length of the observation period.
The $p=0.01$ curve in Fig.~\ref{fig:resolution_vs_dt} is only slightly higher than the $p=0.05$ curve, implying that a confidence level above the 95\% threshold may be reasonably achievable for many deployments.

We can see in \cref{fig:resolution_vs_efficiency,fig:resolution_vs_bg,fig:resolution_vs_dt} that lower background, higher efficiency, and longer observation times lead to a higher discriminatory power.
We find that increasing efficiency leads to diminishing returns, while the effect of reducing background is linear.
However, this trade-off is deployment specific---the slopes of the lines in Fig.~\ref{fig:resolution_vs_bg} depend on the declared MPF (steeper for larger MPFs). Also, practical background reduction techniques may have higher order effects, such as the institution of more restrictive energy cuts leading to a lower neutrino detection rate.

\section{Conclusion}

We have presented a simple observable and associated statistical test to determine possible plutonium diversion from fuel intended for nuclear reactor operation. The method relies on measuring the fraction of MOX fuel present in a reactor using an antineutrino detector near a commercial power reactor partially loaded with MOX fuel.
To this end, we generated time-dependent, isotope-specific fission rates using the Serpent neutronics code base to model pin cells with different fuels and used a weighted sum to determine the fission rates for an entire reactor with a given fraction of MOX fuel.
We then compared reactors with different initial fuel loadings using the straightforward statistical test we developed.

The presented statistical test contains information not only on the absolute rate of neutrino observations but also on the neutrino count rate evolution.
Both of these attributes of the reactor neutrino emissions are useful in characterizing the amount of plutonium in a reactor due to the differences in the neutrino emissions of the four primary fissioning isotopes and the difference in fuel evolution between MOX fuel and LEU fuel.

The presented statistical test considers both the Monte Carlo and the cross-section uncertainty.
In optimal conditions, \textit{i.e.,} zero background, 100\% detector efficiency, and 540-day observation period, this method is capable of signaling the removal of as little as 80~kg of weapons-grade plutonium with a detector containing 5~tons of liquid scintillator placed 25~m from the reactor core.

The IAEA has a stated goal to be able to determine the diversion of a single significant quantity (8~kg of plutonium containing less than 80\% of $^{238}$Pu) of plutonium in 90~days.
The method as described here does not reach that ultimate goal but has demonstrated improved sensitivity in some cases compared to prior work~\cite{Huber_2017, Bernstein_2018}.
In the future we wish to combine our new statistical test with other techniques, notably spectral analysis, to compound the sensitivity of MOX diversion detection and achieve the IAEA target.

In addition to exploring the capabilities of this statistical test, we also found insignificant changes by employing any of the various neutrino spectral models~\cite{neutrino2022,huber2011determination, mueller2011improved, Hayen_2019, Estienne_2019}.
 
Furthermore, as one would expect, decreasing detector efficiency, increasing background rates, and decreasing the duration of the observation period lessens the ability of our statistical test to distinguish between reactors with different fuel loading and thus the diversion of fresh MOX fuel and WGPu.

We have demonstrated that this rate-based neutrino measurement has applications for estimating reactor fuel composition and can be useful within a semi-cooperative framework to determine anomalous activity.
One hurdle for improving our method is a better determination of the time-dependent, isotope-specific fission rates. Any improvements would need to be done using a high-fidelity neutronics model that factors in the complexity of an assembly-level simulation as well as the complexity of fuel shuffling throughout the fuel cycle.
Recent efforts towards addressing this challenge have been published in Ref.~\cite{Price:2024}.
Despite the challenges such a measurement has, the method adds another demonstration of the potential utility of reactor safeguarding using neutrino detectors.

\section{Acknowledgments}
This work was partially supported by the Department of Energy National Nuclear Security Administration, Consortium for Monitoring, Verification and Technology (DE-NE000863). JW would like to thank N.S. Bowden for discussions and the Center for Global Security Research (CGSR) at Lawrence Livermore Laboratory for residency during parts of the research of this publication.
Also, this material is based on work supported under a Department of Energy, Office of Nuclear Energy, Integrated University Program Graduate Fellowship.

\begin{spacing}{1}
\bibliographystyle{bibformat}
\bibliographystyle{unsrt}
\bibliography{references}
\end{spacing}

\end{document}